\documentclass[prl,aps,floatfix,twocolumn,amsmath,amssymb,showpacs]{revtex4}

\usepackage[pdftex]{graphicx}
\usepackage{color}
\usepackage{amsmath}
\usepackage{bm}
\usepackage{times}

\def\nn{\nonumber\\}
\def\vt#1{\ensuremath{\mathbf{#1}}}

\newcommand{\vect}[1]{\mathbf{#1}}
\newcommand{\abs}[1]{\ensuremath{\lvert #1 \rvert}}

\begin{document}
\title{Non-Fermi liquid fixed point for an imbalanced gas of fermions in $1+\epsilon$ dimensions}
\date{\today}
\author{A. J. A. James and A. Lamacraft}
\affiliation{Department of Physics, University of Virginia,
  Charlottesville, VA 22904-4717, USA}
\pacs{}
\begin{abstract}
We consider a gas of two species of fermions with population imbalance. Using the renormalization group in $d=1+\epsilon$ dimensions, we show that for spinless fermions and $\epsilon > 0$ a fixed point appears at finite attractive coupling where the quasiparticle residue vanishes, and identify this with the transition to Larkin--Ovchinnikov--Fulde--Ferrell order (inhomogeneous superconductivity). When the two species of fermions also carry spin degrees of freedom we find a fixed point indicating a transition to spin density wave order.
\end{abstract}
\maketitle

Experiments on ultracold atomic gases allow fermionic pairing phenomena to be investigated with a degree of control and purity hitherto unknown in solid state systems. The preeminent example is the observation of the crossover from Bose-Einstein condensation (BEC) to Bardeen--Cooper--Schreiffer (BCS) superfluidity effected by tuning the scattering length between two atomic components using a Feshbach resonance~\cite{regal2004,zwierlein2004,kinast2004,chin2004}.

It is standard lore that pairing between two components at equal densities -- hence with equal Fermi wavevectors -- occurs for arbitrarily weak interactions in the ground state, though the transition temperature may become very small. By contrast an imbalance in the two populations requires a sufficiently strong attraction before pairing takes place~\cite{Sarma1963}. The first experiments on the imbalanced system~\cite{Partridge2006,Zwierlein2006,Partridge2006a,Shin2006} confirmed this picture, together with the expectation that the transition between the normal and BEC superfluid states is first order, leading to phase separation. 

\begin{figure}
\includegraphics[width=0.4\textwidth]{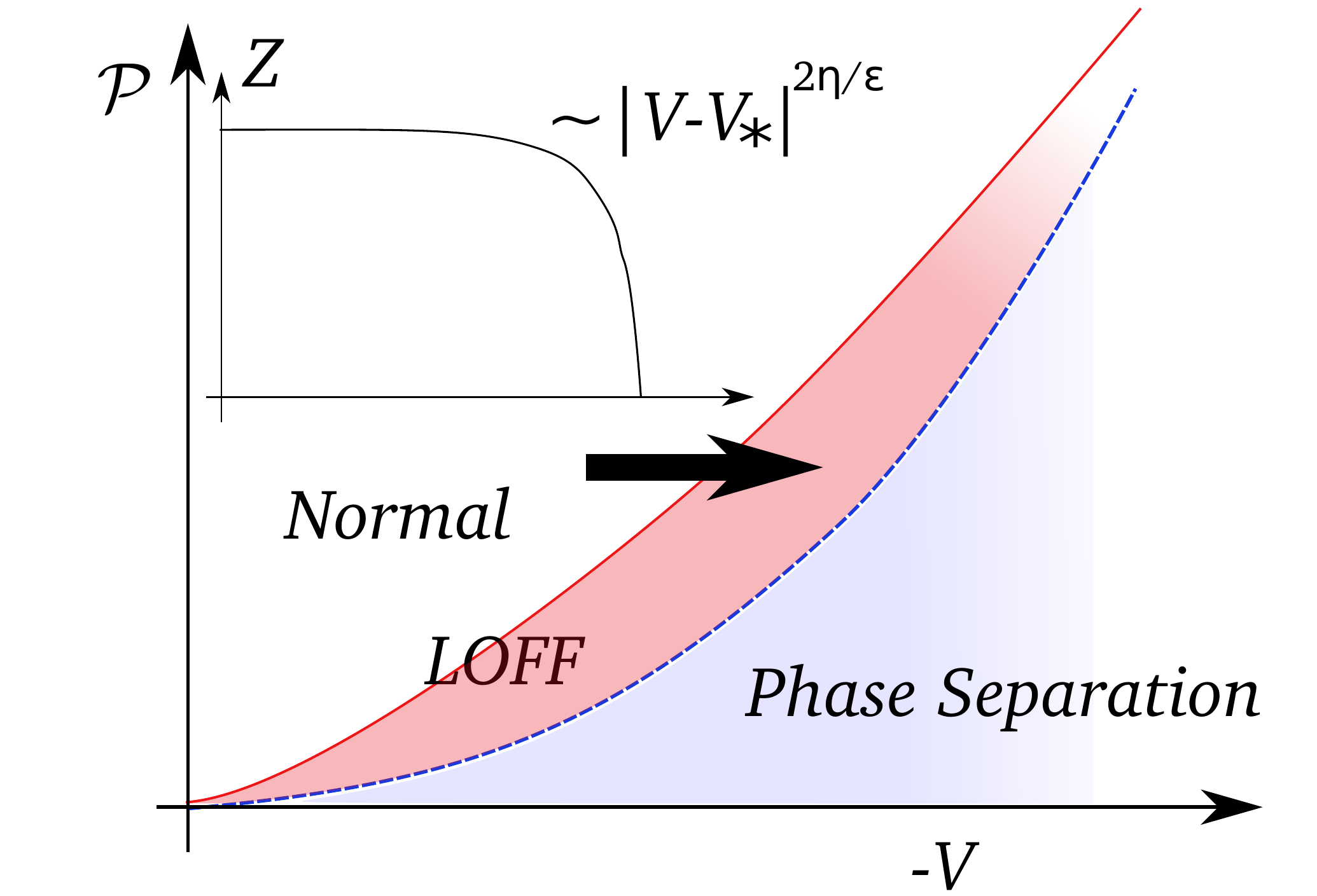}
\caption{(Colour online) Schematic phase diagram showing the continuous quantum phase transition between the normal Fermi liquid and LOFF states as a function of interaction strength and imbalance. The LOFF--BCS transition is first order, leading to phase separation. Inset: vanishing of the quasiparticle residue in the Fermi liquid as the transition is approached along the bold horizontal arrow. Note that the exponent varies continuously with $\mathcal{P}$.
\label{fig:schem}}
\end{figure}

An alternative route to pairing in the imbalanced case was introduced in Refs.~\cite{fulde1964superconductivity,larkin1965inhomogeneous}. The Larkin--Ovchinnikov--Fulde--Ferrell (LOFF) state, as it is now called, is formed from pairs with center--of--mass momentum equal to the difference in Fermi momenta. Within mean--field theory the LOFF phase occupies a rather small part of the phase diagram in terms of imbalance and interaction strength~\cite{Sheehy2007} (see Fig.~\ref{fig:schem}). In lower spatial dimension, the LOFF phase is expected to be more prominent~\cite{lessthan3D}. Experimentally, there is some evidence that the LOFF state occurs in the heavy fermion superconductor CeCoIn$_5$~\cite{radovan2003magnetic,bianchi2003possible}.

In distinction to the normal--BCS transition, the normal--LOFF transition is expected to be continuous at low temperatures. It is therefore somewhat surprising that to date there has been no attempt to understand the nature of the quantum phase transition out of the Fermi liquid state that occurs with increasingly attractive interaction. It is the purpose of this Letter to provide that understanding.

Our main findings can be summarized as follows. Along the line that forms the phase boundary between the normal and LOFF states in a diagram of polarization versus interaction strength (solid red line in Fig.~\ref{fig:schem}) the critical state is characterized by a singular interaction between fermions of the two species with anti--parallel momenta. This leads to an incoherent spectral function for particles at the Fermi surface 
\begin{align}\label{incoh}
A_s(|\vect{K}|=K_{F,s},\omega) \sim \omega^{\eta-1},
\end{align}
where $s=a,b$ labels the species, 
with an exponent $\eta$ that varies continuously along the critical line as a function of the polarization $\mathcal{P}=(n_a-n_b)/(n_a+n_b)$. This line is a higher dimensional analog of the Luttinger liquid in one dimension. For weak polarization, $\mathcal{P} \ll 1$, this result is valid for spatial dimension $1< d \le 2$. As the transition is approached from the normal phase, the quasiparticle residue vanishes continuously. It is our hope that some subset of these predictions can be probed in an ultracold gas by the recently developed technique of momentum--resolved RF spectroscopy~\cite{dao2007,stewart2008}. Towards the end of this work we present an extension to the problem of spin density wave ordering in systems with particle and hole Fermi surfaces, which may of relevance to the recently discovered iron--based pnictide superconductors~\cite{kamihara2008}.

In what follows we assume $n_a > n_b$ and hence the Fermi wave vectors obey $K_{F,a} >K_{F,b}$. 
The energy of a fermion of species $s$ is given by $\xi_{s}(\vt{K}) = \epsilon_{s}(\vt{K})-\mu_{s},$ where $\epsilon_s(\vect{K})=\vect{K}^2/2m_s$ and $\mu_s=K_{F,s}^2/2m_s$, with mass $m_{s}$.
Near their respective Fermi surfaces, where we expect the important physics to occur, the fermions have an approximately linear dispersion: $\xi_{s}(\vt{K})  =v_{F,s} k +\mathcal{O}(k^2)$, where $v_{F,s} = K_{F,s}/m_{s}$ is the Fermi velocity and $k= K-K_{F,s}$ is the momentum relative to the surface.
The effective fermionic Hamiltonian we will work with is
\begin{multline}
\mathcal{H}=\sum_{\vect{K},s}\xi_{s}(\vect{K})\psi^\dagger_{s}(\vect{K})\psi_{s}(\vect{K})\\
+V\sum_{\vect{KK}'\vect{Q}}\psi^\dagger_{b}(\vect{Q}-\vect{K})\psi^\dagger_{a}(\vect{K})\psi_{a}(\vect{K}')\psi_{b}(\vect{Q}-\vect{K}'),
\end{multline}
with a point interaction $V$ that acts only between the different components and where we have set $\hbar=1$.
We start by motivating the problem of pairing with unequal Fermi surfaces.

\begin{figure}
\includegraphics[width=0.3\textwidth]{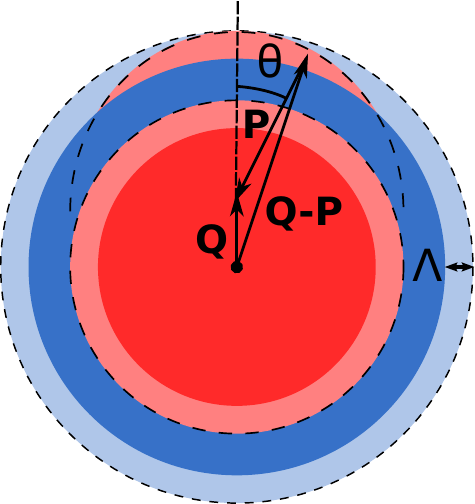}
\caption{(Colour online) Pairing of majority (blue) and minority (red) fermions at fixed total momentum $\vect{Q}$.
\label{fig:shell}}
\end{figure}
\emph{Cooper's problem} Consider a pair of fermions of either species, above their respective Fermi seas. The two fermions interact only with each other, with the Fermi seas serving only to block states below the Fermi level~\cite{cooper1956}. In vacuum the two--particle Schr\"odinger equation takes the form
\begin{equation}\label{2p_sch}
\left[-\frac{\nabla_a^2}{2m_a}-\frac{\nabla_b^2}{2m_b}+V\delta(\vect{r}_a-\vect{r}_b)-E\right]\psi(\vect{r}_a,\vect{r}_b)=0
\end{equation}
After imposing the restrictions due to Pauli blocking, Eq.~(\ref{2p_sch}) is equivalent to the condition
\begin{align}\label{pp_int}
-\frac{1}{V}= \mathop{\int_{|\vect{Q}-\vect{p}|>K_{F,a}}}_{|\vect{p}|>K_{F,b}} \frac{d^d p}{\left(2\pi\right)^2}\frac{1}{\epsilon_a(\vect{Q}-\vect{p})+\epsilon_b(\vect{p})-E} .
\end{align}
where $\vect{Q}$ is the center--of--mass momentum. A bound state corresponds to $E=\mu_a+\mu_b+E_b$ for some $E_b<0$, and to find such a solution for $V$ small requires that the momenta of the two particles are close to antiparallel. Setting $|\vect{Q}|=K_{F,a}-K_{F,b}$ and imposing a momentum shell cut--off of thickness $\Lambda$ around each Fermi surface, we see that the angle between $\vect{p}$ and $\vect{Q}$ is limited as $p-K_{F,b}\to 0$ by the condition (see Fig.~\ref{fig:shell})
\[\theta<\sqrt{\frac{2K_{F,a}(p-K_{F,b})}{Q K_{F,b}}}\]
%
Setting $E_b=0$ to find the threshold for bound state formation,  the right hand side of Eq.~(\ref{pp_int}) becomes for $\Lambda\ll K_{F,s}$
\begin{widetext}
\begin{align}\label{cooper_shell2}
\frac{S_{d-1}}{(2\pi)^d(d-1)}\left(\frac{2K_{F,a}K_{F,b}}{Q}\right)^{(d-1)/2}\int_{K_{F,b}}^{K_{F,b}+\Lambda}dp \frac{(p-K_{F,b})^{(d-1)/2}}{2\bar{v}_F\left(p-K_{F,b}\right)}=\frac{2S_{\epsilon}}{(2\pi)^d\epsilon^2}\left(\frac{2K_{F,a}K_{F,b}}{Q}\right)^{\epsilon/2}\frac{\Lambda^{\epsilon/2}}{2\bar{v}_F}
\end{align}
\end{widetext}
where $S_{d}=2\pi^{d/2}/\Gamma(d/2)$ is the area of the unit sphere in $d$ dimensions and $\bar{v}_F=(v_{F,a}+v_{F,b})/2$.
For small $\epsilon\equiv d-1$, $S_\epsilon\sim \epsilon$, Eq.~(\ref{pp_int}) becomes
\begin{align}\label{cooper_shell2}
-\frac{1}{V}=\frac{1}{2\pi\bar{v}_{F}}\left(\frac{2K_{F,a}K_{F,b}}{Q}\right)^{\epsilon/2}\frac{\Lambda^{\epsilon/2}-\mu^{\epsilon/2}}{\epsilon},
\end{align}
where we have introduced a small infrared cutoff $\mu$. In the limit $\epsilon \to 0$ there is a logarithmic singularity $\log(\Lambda/\mu)$, leading to a logarithmically small bound state energy for arbitrary negative $V$. This divergence in the particle--particle (PP) scattering channel is a consequence of Fermi surface nesting, $\xi_a(\vt{Q}-\vt{p})+\xi_b(\vt{p})=0$, for antiparallel fermions of the two species, familiar from the usual Cooper problem and the BCS theory. The feature that we wish to emphasize is that in the presence of imbalanced Fermi surfaces, the interaction of antiparallel fermions only produces a logarithm in $d=1$. In the renormalization group (RG) sense it is marginal in $d=1$, becoming irrelevant for $d>1$. The above calculation neglects however the particle--hole (PH) contribution to scattering, which  motivates the following more careful RG analysis.

\emph{RG Calculation}
\begin{figure}[t]
\includegraphics[width=0.4\textwidth]{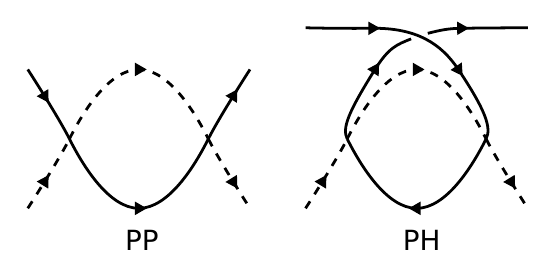}
\caption{The two diagrams contributing to the vertex at one loop order. Dashed and solid lines indicate impurity (a) and gas (b) propagators respectively. The notation is explained in the text.
\label{fig:diag2}}
\end{figure}
The scattering behaviour is encapsulated by the four point vertex function, $\Gamma$, the only contributions to which at one loop order are the bubble diagrams shown in Fig. \ref{fig:diag2}, corresponding to PP and PH excitations. For pair momentum $Q=K_{F,a}-K_{F,b}$ and vanishing external frequency, $\omega \to 0$, these may be evaluated for small angles. The result for the PP bubble is
\begin{align}
-V^2\frac{4 S_{\epsilon}}{(2\pi)^d \epsilon^2} \Big( \frac{2 K_{F,b} K_{F,a}}{Q}  \Big)^{\epsilon/2} \frac{\Lambda^{\epsilon/2}}{2\bar{v}_F}.
\label{eqnpponeloop}
\end{align}
Here the factor of two relative to Eq. (\ref{cooper_shell2}) is due to an equal contribution from pairing below the Fermi surfaces. A comparable treatment results in a similar expression for the PH bubble, but with the opposite sign and $Q \to Q'=K_{F,a}+K_{F,b}$.
Combining the terms, we have for scattering of an antiparallel pair at the Fermi surface
\begin{align}
\label{eqngamma}
\Gamma=V-V^2\frac{2S_{\epsilon} (2\sqrt{K_{F,b}K_{F,a}})^{\epsilon/2}}{(2\pi)^{1+\epsilon} \bar{v}_F\epsilon^2}\Lambda^{\epsilon/2}F+\hdots,
\end{align}
with 
\begin{equation}\label{eqnF}
F=\frac{Q^{-\epsilon/2}-Q'^{-\epsilon/2}}{(K_{F,a}K_{F,b})^{-\epsilon/4}}.
\end{equation}
%
%
We see that there is cancellation of the two channels if $d=1$ or $K_{F,b}/K_{F,a}=0$, so that $F=0$. 

By demanding cut--off independence, $\Lambda\frac{d}{d \Lambda}\Gamma=0$ and defining a dimensionless coupling, $g$, via
\begin{align}
g = \frac{S_\epsilon(2\Lambda\sqrt{K_{F,b}K_{F,a}})^{\epsilon/2}}{(2\pi)^{1+\epsilon}\epsilon \bar{v}_F } V,
\end{align}
we obtain the beta function
\begin{align}
\beta(g)=-\Lambda\frac{d}{d \Lambda} g = -\frac{\epsilon}{2} g-F g^2+\hdots.
\end{align}
Hence there is a non--trivial fixed point
$g_\star = -\frac{\epsilon}{2F}$,
%
\begin{figure}[t]
\includegraphics[width=0.4\textwidth]{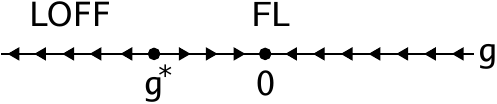}
\caption{Flow diagram for the spinless case, indicating Fermi liquid (FL) and LOFF phases.
\label{fig:spinlessflow}}
\end{figure}
which is unstable as illustrated in Fig.\ \ref{fig:spinlessflow}.
Notice the difference from the more familiar situation, typified by the Wilson--Fisher fixed point, in which the interacting and free fixed points merge as the critical dimension is approached. In this case, the vanishing of $F$ as $\epsilon\to 0$ means that $\beta$ vanishes, leaving the fixed point at finite coupling. The $\epsilon=0$ cancellation extends to all orders (this may be shown, for example, using Ward identities~\cite{metzner1993conservation}) and hence $\beta(g)=0$ for a range of $g$, indicating the existence of the well known Luttinger liquid critical phase.

For $\epsilon>0$ and weak imbalance, $K_{F,a} \sim K_{F,b}$,  the PP contribution to Eq. (\ref{eqnF}) dominates and the fixed point occurs at weak coupling
\begin{align}
g_\star=-\frac{\epsilon}{2}\Big(\frac{K_{F,a}-K_{F,b}}{\sqrt{K_{F,a}K_{F,a}}}\Big)^{\epsilon/2}.
\end{align}
Integrating the flow for $\abs{g}\ll \abs{g_\star}$ yields $g= g_0 (\Lambda/\Lambda_0)^{\epsilon/2}$, where $g_0$ and $\Lambda_0$ are the initial conditions for the scaling. Consequently $V$ is independent of $\Lambda$. This is the expected result for couplings when $\mathcal{P}=0$ and $Q\ne 0$ \cite{shankar1994renormalization}. In contrast, when $g=g_\star$ we find $V\sim -\Lambda^{-\epsilon/2}$ so that the coupling becomes increasingly attractive as $\Lambda$ decreases.
The situation for $\mathcal{P}\ll 1$ corresponds to a microscopic perturbation theory in which the vertex is given as a geometric sum of PP bubbles and the resulting expression can be extended to $\epsilon=1$. With this in mind we expect the weak polarization fixed point result to be valid in $d=2$.

For strong imbalance we may expand $F$ as a power series in $K_{F,b}/K_{F,a}$, $F \approx \epsilon K_{F,b}/K_{F,a}$. As a result $\beta$ is given by
\begin{align}
\beta(g)=-\frac{\epsilon}{2} g-\epsilon \frac{K_{F,b}}{K_{F,a}} g^2+\hdots.
\end{align}
The fixed point is now at strong coupling, independent of $\epsilon$, and higher order contributions should then be taken into account.
As such it is not possible to make quantitative statements about the strongly imbalanced case, $\mathcal{P} \sim 1$ using this method.

We now turn to the fermion self energy. The one loop contribution produces a routine frequency independent shift which may be absorbed by the chemical potential, so as to keep $K_{F,s}$ fixed. The first significant behaviour is found by calculating the two loop `rising sun' diagram, yielding 
\begin{align}\label{se}
\frac{\partial\mathrm{Re}\,\Sigma^R_s(|\vect{K}|=K_{F,s},\omega)}{\partial\omega}\Big|_{\omega=0} = \frac{g^2C_{s,\epsilon}}{4\epsilon}  +\mathcal{O}(g^3).
\end{align}
Though the coefficient $C_{s,\epsilon}$ depends on the ratio $K_{F,a}/K_{F,b}$, it goes to unity in the limit $\epsilon\to 0$. Note that Eq.~(\ref{se}) arises from scattering of almost antiparallel particles of the two species.
 
 $\Lambda$ independence of the physical correlation functions necessitates the introduction of a multiplicative field renormalization $\psi_{s}^{\mathrm{phys}}=Z^{1/2}\psi_s$. $Z$ is fixed by demanding $\Lambda$ independence of the physical Greens function 
\begin{align}
G^{\mathrm{phys}}_s(|\vect{K}|=K_{F,s},\omega)=\frac{Z}{\omega-\Sigma^R_s(|\vect{K}|=K_{F,s},\omega)}.
\end{align}
This implies
\begin{align}
 \frac{d\ln Z}{d\ln \Lambda}=-\frac{g}{2\epsilon}\beta(g)=\frac{g^2}{4}+\mathcal{O}(g^3),
\end{align}
so that at $g=g_\star$, $Z \to 0$ as $\Lambda\to 0$. At the fixed point the scaling behaviour of the propagator is given by
\begin{align}
G^{\mathrm{phys}}_s(|\vect{K}|=K_{F,s},\omega) \sim \omega^{\eta-1},
\end{align}
where the critical exponent $\eta=g_\star^2/4=\epsilon^2/16F^2$. Hence the quasiparticle pole is replaced by a branch cut with attendant power law divergent spectral function. It is also enlightening to determine the dependence of $Z$ on the coupling as it approaches $g_\star$. By linearizing the flow equations near the fixed point
we may derive the following relation between the start and end points of the flow, $(Z_0,g_0)$ and $(Z,g)$ respectively,
\begin{align}
Z&=Z_0\Big(\frac{g_0-g_\star}{g-g_\star}\Big)^{\epsilon/8F^2}\exp\Big[-\frac{2}{F} (g-g_0)\Big].
\end{align}
Using the above we may extract the value of the Fermi liquid quasiparticle residue, $Z$, due to an initial coupling $g_0$. If we take $g_0 > g_\star$ so that the coupling flows to $g=0$, the final value of $Z$ will depend on $g_0$ as $Z \sim (g_0-g_\star) ^{\epsilon/8F^2}$ (see inset to Fig.~\ref{fig:schem}). 

It is possible to extend the calculation above to treat components that themselves have internal degrees of freedom. In the case that $a$ and $b$ are both spin half particles, the quartic part of the Hamiltonian becomes
\begin{align}
\label{eqnspinhamiltonian}
\mathcal{H}_{int}=\sum_{\sigma,\sigma'}\big[&(V_{e\parallel}\delta_{\sigma,\sigma'}+V_{e\perp}\delta_{\sigma,-\sigma'}  )\psi^\dagger_{b,\sigma}\psi^\dagger_{a,\sigma'}\psi_{a,\sigma}\psi_{b,\sigma'} \nn
+&(V_{d\parallel}\delta_{\sigma,\sigma'}+V_{d\perp}\delta_{\sigma,-\sigma'}  ) \psi^\dagger_{b,\sigma}\psi^\dagger_{a,\sigma'}\psi_{a,\sigma'}\psi_{b,\sigma}\big].
\end{align}
where we have supressed momentum labels.
Clearly there is no discernible difference between processes due to $V_{e\parallel}$ and those due to $V_{d\parallel}$. Therefore we define $V_\parallel=V_{e\parallel}+V_{d\parallel}$. If $V_\parallel=-V_{d\perp}$ Eq. (\ref{eqnspinhamiltonian}) is equivalent to an anisotropic Heisenberg interaction, with $J_z=4V_{\parallel}$ and $J_{xy}=2V_{e\perp}$.
Keeping only the lowest order terms (so that $F=0$) and defining dimensionless couplings, $g_l=V_lA\epsilon$, where $A=(K_{F,b}\Lambda)^{\epsilon/2}/\pi\bar{v}_F\epsilon$, one finds
\begin{align}
\beta(g_\parallel)&=-\frac{\epsilon}{2}g_\parallel+\frac{1}{2}g_{e\perp}^2,\label{eqnflowparallel}\\
\beta(g_{d\perp})&=-\frac{\epsilon}{2}g_{d\perp}-\frac{1}{2}g_{e\perp}^2,\label{eqnflowdperp}\\
\beta(g_{e\perp})&=-\frac{\epsilon}{2}g_{e\perp}-g_{e\perp}(g_{d\perp}-g_\parallel) \label{eqnfloweperp}.
\end{align}
The first two equations are equivalent if $V_\parallel=-V_{d\perp}$. Focusing on this case we discern that there is a non--trivial fixed point at $(g_{\parallel\star},g_{e\perp \star})=(\epsilon/4,\epsilon/2)$. Concomitantly, and in terms of the Heisenberg exchange couplings, there is a transition to spin density wave order at antiferromagnetic exchange $J_{z\star}=J_{xy\star}=\epsilon \pi \bar{v}_F/(K_{F,b} \Lambda)^{\epsilon/2}$.

An experimentally relevant scenario is that of interacting electron and hole pockets~\cite{pieri2007}. Such systems have been investigated extensively in the context of excitonic transitions  and antiferromagnetism (see for example ~\cite{rice1970}) and recently because of interest in the high temperature superconductivity of iron pnictides \cite{chubukov2009}. In the calculation presented here, switching one of the components from particle--like to hole--like alters the dispersion as $\xi_{\vt{K},s}\to\xi_{\vt{K},h}=-v_{F,h} k$ and changes the pairing wavevector to $\vert\vt{Q}\rvert=K_{F,e}+K_{F,h}$ where the subscripts $e$ and $h$ indicate electrons and holes respectively. In terms of Eqs. (\ref{eqnflowparallel}--\ref{eqnfloweperp}) the effect is to flip the sign of the $\mathcal{O}(g^2)$ terms. Again taking $V_\parallel=-V_{d\perp}$, the fixed point is at $(g_{\parallel\star},g_{e\perp\star})=(-\epsilon/4,\epsilon/2)$ or equivalently, for ferromagnetic exchange $J_{z\star}=-J_{xy\star}=-\epsilon \pi \bar{v}_F/(K_{F,b} \Lambda)^{\epsilon/2}$.

In this letter we have applied the renormalization group to polarized two species fermi gases in $1+\epsilon$ dimensions. The central result for spinless fermions is the appearance of a non--Fermi liquid fixed point, characterised by finite pairing wave vector (centre of mass momentum) and power law spectral function, with an exponent that depends on the polarization. For fermions that carry spin we have shown that there are non--trivial fixed points that describe magnetic ordering of the spin density wave type. AL gratefully acknowledges the support of the NSF under grant DMR-0846788.

\begin{thebibliography}{27}
\expandafter\ifx\csname natexlab\endcsname\relax\def\natexlab#1{#1}\fi
\expandafter\ifx\csname bibnamefont\endcsname\relax
  \def\bibnamefont#1{#1}\fi
\expandafter\ifx\csname bibfnamefont\endcsname\relax
  \def\bibfnamefont#1{#1}\fi
\expandafter\ifx\csname citenamefont\endcsname\relax
  \def\citenamefont#1{#1}\fi
\expandafter\ifx\csname url\endcsname\relax
  \def\url#1{\texttt{#1}}\fi
\expandafter\ifx\csname urlprefix\endcsname\relax\def\urlprefix{URL }\fi
\providecommand{\bibinfo}[2]{#2}
\providecommand{\eprint}[2][]{\url{#2}}

\bibitem[{\citenamefont{Regal et~al.}(2004)\citenamefont{Regal, Greiner, and
  Jin}}]{regal2004}
\bibinfo{author}{\bibfnamefont{C.~A.} \bibnamefont{Regal}},
  \bibinfo{author}{\bibfnamefont{M.}~\bibnamefont{Greiner}}, \bibnamefont{and}
  \bibinfo{author}{\bibfnamefont{D.~S.} \bibnamefont{Jin}},
  \bibinfo{journal}{Phys. Rev. Lett.} \textbf{\bibinfo{volume}{92}},
  \bibinfo{pages}{040403} (\bibinfo{year}{2004}).

\bibitem[{\citenamefont{Zwierlein et~al.}(2004)\citenamefont{Zwierlein, Stan,
  Schunck, Raupach, Kerman, and Ketterle}}]{zwierlein2004}
\bibinfo{author}{\bibfnamefont{M.~W.} \bibnamefont{Zwierlein}},
  \bibinfo{author}{\bibfnamefont{C.~A.} \bibnamefont{Stan}},
  \bibinfo{author}{\bibfnamefont{C.~H.} \bibnamefont{Schunck}},
  \bibinfo{author}{\bibfnamefont{S.~M.~F.} \bibnamefont{Raupach}},
  \bibinfo{author}{\bibfnamefont{A.~J.} \bibnamefont{Kerman}},
  \bibnamefont{and} \bibinfo{author}{\bibfnamefont{W.}~\bibnamefont{Ketterle}},
  \bibinfo{journal}{Phys.\ Rev.\ Lett.} \textbf{\bibinfo{volume}{92}},
  \bibinfo{pages}{120403} (\bibinfo{year}{2004}).

\bibitem[{\citenamefont{Kinast et~al.}(2004)\citenamefont{Kinast, Hemmer, Gehm,
  Turlapov, and Thomas}}]{kinast2004}
\bibinfo{author}{\bibfnamefont{J.}~\bibnamefont{Kinast}},
  \bibinfo{author}{\bibfnamefont{S.~L.} \bibnamefont{Hemmer}},
  \bibinfo{author}{\bibfnamefont{M.~E.} \bibnamefont{Gehm}},
  \bibinfo{author}{\bibfnamefont{A.}~\bibnamefont{Turlapov}}, \bibnamefont{and}
  \bibinfo{author}{\bibfnamefont{J.~E.} \bibnamefont{Thomas}},
  \bibinfo{journal}{Phys. Rev. Lett.} \textbf{\bibinfo{volume}{92}},
  \bibinfo{pages}{150402} (\bibinfo{year}{2004}).

\bibitem[{\citenamefont{Chin et~al.}(2004)\citenamefont{Chin, Bartenstein,
  Altmeyer, Riedl, Jochim, Denschlag, and Grimm}}]{chin2004}
\bibinfo{author}{\bibfnamefont{C.}~\bibnamefont{Chin}},
  \bibinfo{author}{\bibfnamefont{M.}~\bibnamefont{Bartenstein}},
  \bibinfo{author}{\bibfnamefont{A.}~\bibnamefont{Altmeyer}},
  \bibinfo{author}{\bibfnamefont{S.}~\bibnamefont{Riedl}},
  \bibinfo{author}{\bibfnamefont{S.}~\bibnamefont{Jochim}},
  \bibinfo{author}{\bibfnamefont{J.~H.} \bibnamefont{Denschlag}},
  \bibnamefont{and} \bibinfo{author}{\bibfnamefont{R.}~\bibnamefont{Grimm}},
  \bibinfo{journal}{Science} \textbf{\bibinfo{volume}{305}},
  \bibinfo{pages}{1128} (\bibinfo{year}{2004}).

\bibitem[{\citenamefont{Sarma}(1963)}]{Sarma1963}
\bibinfo{author}{\bibfnamefont{G.}~\bibnamefont{Sarma}}, \bibinfo{journal}{J.
  Phys. Chem. Solids} \textbf{\bibinfo{volume}{24}}, \bibinfo{pages}{1029}
  (\bibinfo{year}{1963}).

\bibitem[{\citenamefont{Partridge
  et~al.}(2006{\natexlab{a}})\citenamefont{Partridge, Li, Kamar, Liao, and
  Hulet}}]{Partridge2006}
\bibinfo{author}{\bibfnamefont{G.}~\bibnamefont{Partridge}},
  \bibinfo{author}{\bibfnamefont{W.}~\bibnamefont{Li}},
  \bibinfo{author}{\bibfnamefont{R.}~\bibnamefont{Kamar}},
  \bibinfo{author}{\bibfnamefont{Y.}~\bibnamefont{Liao}}, \bibnamefont{and}
  \bibinfo{author}{\bibfnamefont{R.}~\bibnamefont{Hulet}},
  \bibinfo{journal}{Science} \textbf{\bibinfo{volume}{311}},
  \bibinfo{pages}{503} (\bibinfo{year}{2006}{\natexlab{a}}).

\bibitem[{\citenamefont{Zwierlein et~al.}(2006)\citenamefont{Zwierlein,
  Schirotzek, Schunck, and Ketterle}}]{Zwierlein2006}
\bibinfo{author}{\bibfnamefont{M.}~\bibnamefont{Zwierlein}},
  \bibinfo{author}{\bibfnamefont{A.}~\bibnamefont{Schirotzek}},
  \bibinfo{author}{\bibfnamefont{C.}~\bibnamefont{Schunck}}, \bibnamefont{and}
  \bibinfo{author}{\bibfnamefont{W.}~\bibnamefont{Ketterle}},
  \bibinfo{journal}{Science} \textbf{\bibinfo{volume}{311}},
  \bibinfo{pages}{492} (\bibinfo{year}{2006}).

\bibitem[{\citenamefont{Partridge
  et~al.}(2006{\natexlab{b}})\citenamefont{Partridge, Li, Liao, Hulet, Haque,
  and Stoof}}]{Partridge2006a}
\bibinfo{author}{\bibfnamefont{G.}~\bibnamefont{Partridge}},
  \bibinfo{author}{\bibfnamefont{W.}~\bibnamefont{Li}},
  \bibinfo{author}{\bibfnamefont{Y.}~\bibnamefont{Liao}},
  \bibinfo{author}{\bibfnamefont{R.}~\bibnamefont{Hulet}},
  \bibinfo{author}{\bibfnamefont{M.}~\bibnamefont{Haque}}, \bibnamefont{and}
  \bibinfo{author}{\bibfnamefont{H.}~\bibnamefont{Stoof}},
  \bibinfo{journal}{Phys. Rev. Lett.} \textbf{\bibinfo{volume}{97}},
  \bibinfo{pages}{190407} (\bibinfo{year}{2006}{\natexlab{b}}).

\bibitem[{\citenamefont{Shin et~al.}(2006)\citenamefont{Shin, Zwierlein,
  Schunck, Schirotzek, and Ketterle}}]{Shin2006}
\bibinfo{author}{\bibfnamefont{Y.}~\bibnamefont{Shin}},
  \bibinfo{author}{\bibfnamefont{M.}~\bibnamefont{Zwierlein}},
  \bibinfo{author}{\bibfnamefont{C.}~\bibnamefont{Schunck}},
  \bibinfo{author}{\bibfnamefont{A.}~\bibnamefont{Schirotzek}},
  \bibnamefont{and} \bibinfo{author}{\bibfnamefont{W.}~\bibnamefont{Ketterle}},
  \bibinfo{journal}{Phys. Rev. Lett.} \textbf{\bibinfo{volume}{97}},
  \bibinfo{pages}{30401} (\bibinfo{year}{2006}).

\bibitem[{\citenamefont{Fulde and Ferrell}(1964)}]{fulde1964superconductivity}
\bibinfo{author}{\bibfnamefont{P.}~\bibnamefont{Fulde}} \bibnamefont{and}
  \bibinfo{author}{\bibfnamefont{R.}~\bibnamefont{Ferrell}},
  \bibinfo{journal}{Phys. Rev} \textbf{\bibinfo{volume}{135}},
  \bibinfo{pages}{A550} (\bibinfo{year}{1964}).

\bibitem[{\citenamefont{Larkin and
  Ovchinnikov}(1965)}]{larkin1965inhomogeneous}
\bibinfo{author}{\bibfnamefont{A.}~\bibnamefont{Larkin}} \bibnamefont{and}
  \bibinfo{author}{\bibfnamefont{I.}~\bibnamefont{Ovchinnikov}},
  \bibinfo{journal}{JETP} \textbf{\bibinfo{volume}{20}}, \bibinfo{pages}{762}
  (\bibinfo{year}{1965}).

\bibitem[{\citenamefont{Sheehy and Radzihovsky}(2007)}]{Sheehy2007}
\bibinfo{author}{\bibfnamefont{D.}~\bibnamefont{Sheehy}} \bibnamefont{and}
  \bibinfo{author}{\bibfnamefont{L.}~\bibnamefont{Radzihovsky}},
  \bibinfo{journal}{Ann. Phys.} \textbf{\bibinfo{volume}{322}},
  \bibinfo{pages}{1790} (\bibinfo{year}{2007}).

\bibitem[{\citenamefont{Mora and Combescot}(2004)}]{lessthan3D}
\bibinfo{author}{\bibfnamefont{C.}~\bibnamefont{Mora}} \bibnamefont{and}
  \bibinfo{author}{\bibfnamefont{R.}~\bibnamefont{Combescot}},
  \bibinfo{journal}{Europhys. Lett.)} \textbf{\bibinfo{volume}{66}},
  \bibinfo{pages}{833} (\bibinfo{year}{2004}),
\bibinfo{author}{\bibfnamefont{M.}~\bibnamefont{Parish}},
  \bibinfo{author}{\bibfnamefont{S.}~\bibnamefont{Baur}},
  \bibinfo{author}{\bibfnamefont{E.}~\bibnamefont{Mueller}}, \bibnamefont{and}
  \bibinfo{author}{\bibfnamefont{D.}~\bibnamefont{Huse}},
  \bibinfo{journal}{Phys. Rev. Lett.} \textbf{\bibinfo{volume}{99}},
  \bibinfo{pages}{250403} (\bibinfo{year}{2007}),
\bibinfo{author}{\bibfnamefont{G.}~\bibnamefont{Conduit}},
  \bibinfo{author}{\bibfnamefont{P.}~\bibnamefont{Conlon}}, \bibnamefont{and}
  \bibinfo{author}{\bibfnamefont{B.}~\bibnamefont{Simons}},
  \bibinfo{journal}{Phys. Rev. A} \textbf{\bibinfo{volume}{77}},
  \bibinfo{pages}{53617} (\bibinfo{year}{2008}),
\bibinfo{author}{\bibfnamefont{F.}~\bibnamefont{Heidrich-Meisner}},
  \bibinfo{author}{\bibfnamefont{A.}~\bibnamefont{Feiguin}},
  \bibinfo{author}{\bibfnamefont{U.}~\bibnamefont{Schollw{\"o}ck}},
  \bibnamefont{and} \bibinfo{author}{\bibfnamefont{W.}~\bibnamefont{Zwerger}},
  \bibinfo{journal}{arxiv:0908.3074}.

\bibitem[{\citenamefont{Radovan et~al.}(2003)\citenamefont{Radovan, Fortune,
  Murphy, Hannahs, Palm, Tozer, and Hall}}]{radovan2003magnetic}
\bibinfo{author}{\bibfnamefont{H.}~\bibnamefont{Radovan}},
  \bibinfo{author}{\bibfnamefont{N.}~\bibnamefont{Fortune}},
  \bibinfo{author}{\bibfnamefont{T.}~\bibnamefont{Murphy}},
  \bibinfo{author}{\bibfnamefont{S.}~\bibnamefont{Hannahs}},
  \bibinfo{author}{\bibfnamefont{E.}~\bibnamefont{Palm}},
  \bibinfo{author}{\bibfnamefont{S.}~\bibnamefont{Tozer}}, \bibnamefont{and}
  \bibinfo{author}{\bibfnamefont{D.}~\bibnamefont{Hall}},
  \bibinfo{journal}{Nature} \textbf{\bibinfo{volume}{425}}, \bibinfo{pages}{51}
  (\bibinfo{year}{2003}).

\bibitem[{\citenamefont{Bianchi et~al.}(2003)\citenamefont{Bianchi, Movshovich,
  Capan, Pagliuso, and Sarrao}}]{bianchi2003possible}
\bibinfo{author}{\bibfnamefont{A.}~\bibnamefont{Bianchi}},
  \bibinfo{author}{\bibfnamefont{R.}~\bibnamefont{Movshovich}},
  \bibinfo{author}{\bibfnamefont{C.}~\bibnamefont{Capan}},
  \bibinfo{author}{\bibfnamefont{P.}~\bibnamefont{Pagliuso}}, \bibnamefont{and}
  \bibinfo{author}{\bibfnamefont{J.}~\bibnamefont{Sarrao}},
  \bibinfo{journal}{Phys. Rev. Lett.} \textbf{\bibinfo{volume}{91}},
  \bibinfo{pages}{187004} (\bibinfo{year}{2003}).

\bibitem[{\citenamefont{Dao et~al.}(2007)\citenamefont{Dao, Georges, Dalibard,
  Salomon, and Carusotto}}]{dao2007}
\bibinfo{author}{\bibfnamefont{T.}~\bibnamefont{Dao}},
  \bibinfo{author}{\bibfnamefont{A.}~\bibnamefont{Georges}},
  \bibinfo{author}{\bibfnamefont{J.}~\bibnamefont{Dalibard}},
  \bibinfo{author}{\bibfnamefont{C.}~\bibnamefont{Salomon}}, \bibnamefont{and}
  \bibinfo{author}{\bibfnamefont{I.}~\bibnamefont{Carusotto}},
  \bibinfo{journal}{Phys. Rev. Lett.} \textbf{\bibinfo{volume}{98}},
  \bibinfo{pages}{240402} (\bibinfo{year}{2007}).

\bibitem[{\citenamefont{Stewart et~al.}(2008)\citenamefont{Stewart, Gaebler,
  and Jin}}]{stewart2008}
\bibinfo{author}{\bibfnamefont{J.}~\bibnamefont{Stewart}},
  \bibinfo{author}{\bibfnamefont{J.}~\bibnamefont{Gaebler}}, \bibnamefont{and}
  \bibinfo{author}{\bibfnamefont{D.}~\bibnamefont{Jin}},
  \bibinfo{journal}{Nature} \textbf{\bibinfo{volume}{454}},
  \bibinfo{pages}{744} (\bibinfo{year}{2008}).

\bibitem[{\citenamefont{Kamihara et~al.}(2008)\citenamefont{Kamihara, Watanabe,
  Hirano, Hosono et~al.}}]{kamihara2008}
\bibinfo{author}{\bibfnamefont{Y.}~\bibnamefont{Kamihara}},
  \bibinfo{author}{\bibfnamefont{T.}~\bibnamefont{Watanabe}},
  \bibinfo{author}{\bibfnamefont{M.}~\bibnamefont{Hirano}},
  \bibinfo{author}{\bibfnamefont{H.}~\bibnamefont{Hosono}},
  \bibnamefont{et~al.}, \bibinfo{journal}{J. Am. Chem. Soc}
  \textbf{\bibinfo{volume}{130}}, \bibinfo{pages}{3296} (\bibinfo{year}{2008}).

\bibitem[{\citenamefont{Cooper}(1956)}]{cooper1956}
\bibinfo{author}{\bibfnamefont{L.}~\bibnamefont{Cooper}},
  \bibinfo{journal}{Phys. Rev.} \textbf{\bibinfo{volume}{104}},
  \bibinfo{pages}{1189} (\bibinfo{year}{1956}).

\bibitem[{\citenamefont{Metzner and Di~Castro}(1993)}]{metzner1993conservation}
\bibinfo{author}{\bibfnamefont{W.}~\bibnamefont{Metzner}} \bibnamefont{and}
  \bibinfo{author}{\bibfnamefont{C.}~\bibnamefont{Di~Castro}},
  \bibinfo{journal}{Phys. Rev. B} \textbf{\bibinfo{volume}{47}},
  \bibinfo{pages}{16107} (\bibinfo{year}{1993}).

\bibitem[{\citenamefont{Shankar}(1994)}]{shankar1994renormalization}
\bibinfo{author}{\bibfnamefont{R.}~\bibnamefont{Shankar}},
  \bibinfo{journal}{Rev. Mod. Phys.} \textbf{\bibinfo{volume}{66}},
  \bibinfo{pages}{129} (\bibinfo{year}{1994}).

\bibitem[{\citenamefont{Pieri et~al.}(2007)\citenamefont{Pieri, Neilson, and
  Strinati}}]{pieri2007}
\bibinfo{author}{\bibfnamefont{P.}~\bibnamefont{Pieri}},
  \bibinfo{author}{\bibfnamefont{D.}~\bibnamefont{Neilson}}, \bibnamefont{and}
  \bibinfo{author}{\bibfnamefont{G.}~\bibnamefont{Strinati}},
  \bibinfo{journal}{Phys. Rev. B} \textbf{\bibinfo{volume}{75}},
  \bibinfo{pages}{113301} (\bibinfo{year}{2007}).

\bibitem[{\citenamefont{Rice}(1970)}]{rice1970}
\bibinfo{author}{\bibfnamefont{T.}~\bibnamefont{Rice}}, \bibinfo{journal}{Phys.
  Rev. B} \textbf{\bibinfo{volume}{2}}, \bibinfo{pages}{3619}
  (\bibinfo{year}{1970}).

\bibitem[{\citenamefont{Chubukov}(2009)}]{chubukov2009}
\bibinfo{author}{\bibfnamefont{A.}~\bibnamefont{Chubukov}},
  \bibinfo{journal}{Physica C}  (\bibinfo{year}{2009}).
\end{thebibliography}

\end{document}